\documentclass[letter,UKenglish]{lipics-v2016}

\usepackage{bbm, hyperref, cleveref}
\usepackage{tikz}
\usepackage{thm-restate}

\newcommand{\f}{\displaystyle\frac}
\newcommand{\cd}{\cdot}

\newcommand{\sr}{\sqrt}
\newcommand{\cds}{\cdots}
\newcommand{\lds}{\ldots}

\newcommand{\bs}{\backslash}
\newcommand{\s}{\subseteq}

\newcommand{\BE}{\begin{enumerate}}
\newcommand{\EE}{\end{enumerate}}
\newcommand{\im}{\item}
\newcommand{\BI}{\begin{itemize}}
\newcommand{\EI}{\end{itemize}}

\newcommand{\logn}{\log n}

\newcommand{\inv}{^{-1}}

\newcommand{\N}{\mathbb N}

\newcommand{\e}{\epsilon}
\newcommand{\de}{\delta}

\newcommand{\be}{\beta}
\newcommand{\om}{\omega}
\newcommand{\Om}{\Omega}
\newcommand{\el}{\ell}

\newcommand{\Th}{\Theta}
\newcommand{\m}{\mathcal}
\newcommand{\tO}{\tilde{O}}

\newcommand{\lc}{\lceil}
\newcommand{\rc}{\rceil}

\newcommand{\E}{\mathbb E}

\newcommand{\poly}{\text{poly}}

\newcommand{\lp}{\left(}
\newcommand{\rp}{\right)}
\newcommand{\lb}{\left[}
\newcommand{\rb}{\right]}
\newcommand{\lmt}{\left[\begin{matrix}}
\newcommand{\rmt}{\end{matrix}\right]}

\newcommand{\BT}{\begin{theorem}}
\newcommand{\ET}{\end{theorem}}
\newcommand{\BL}{\begin{lemma}}
\newcommand{\EL}{\end{lemma}}
\newcommand{\BC}{\begin{corollary}}
\newcommand{\EC}{\end{corollary}}
\newcommand{\BCL}{\begin{claim}}
\newcommand{\ECL}{\end{claim}}
\newcommand{\BO}{\begin{observation}}
\newcommand{\EO}{\end{observation}}
\newcommand{\BD}{\begin{definition}}
\newcommand{\ED}{\end{definition}}
\newcommand{\BF}{\begin{fact}}
\newcommand{\EF}{\end{fact}}
\newcommand{\BP}{\begin{proof}}
\newcommand{\EP}{\end{proof}}
\newcommand{\BPS}{\begin{proof}[Proof (Sketch)]}
\newcommand{\EPS}{\end{proof}}

\renewcommand{\paragraph}[1]{\vspace{2mm}\noindent\textbf{#1}\hspace{2mm}}
\newcommand{\defn}[1]{{\textbf{#1}}}
\newcommand{\para}{\paragraph}

\newcommand{\srlogn}{{\sqrt{\log n}}}
\newcommand{\tmix}{{\tau_{\text{mix}}}}

\title{
New Distributed Algorithms in Almost Mixing Time
via Transformations from Parallel Algorithms}

\author[1]{Mohsen Ghaffari}
\author[2]{Jason Li}
\affil[1]{ETH Zurich, 
  \texttt{ghaffari@inf.ethz.ch}}
\affil[2]{Carnegie Mellon University, 
  \texttt{jmli@cs.cmu.edu}}
\authorrunning{M. Ghaffari and J. Li}
\titlerunning{Distributed Algorithms in Almost Mixing Time}

\Copyright{Mohsen Ghaffari and Jason Li}

\subjclass{G.2.2 Graph Theory - Graph Algorithms}
\keywords{Distributed Graph Algorithms, Multi-Commodity Routing, Maxing Time, Random Graphs}

\EventEditors{John Q. Open and Joan R. Access}
\EventNoEds{2}
\EventLongTitle{42nd Conference on Very Important Topics (CVIT 2016)}
\EventShortTitle{CVIT 2016}
\EventAcronym{CVIT}
\EventYear{2016}
\EventDate{December 24--27, 2016}
\EventLocation{Little Whinging, United Kingdom}
\EventLogo{}
\SeriesVolume{42}
\ArticleNo{23}

\begin{document}

\date{}

\maketitle
\begin{abstract} 
We show that many classical optimization problems --- such as $(1\pm\epsilon)$-approximate maximum flow, shortest path, and transshipment --- can be computed in $\tmix(G)\cdot n^{o(1)}$ rounds of distributed message passing, where $\tmix(G)$ is the mixing time of the network graph $G$. This extends the result of Ghaffari et al.\ [PODC'17], whose main result is a distributed MST algorithm in $\tmix(G)\cdot 2^{O(\sqrt{\log n \log\log n})}$ rounds in the CONGEST model, to a much wider class of optimization problems. For many practical networks of interest, e.g., peer-to-peer or overlay network structures, the mixing time $\tmix(G)$ is small, e.g., polylogarithmic. On these networks, our algorithms bypass the $\tilde\Omega(\sqrt n+D)$ lower bound of Das Sarma et al.\ [STOC'11], which applies for worst-case graphs and applies to all of the above optimization problems. For all of the problems except MST, this is the first distributed algorithm which takes $o(\sqrt n)$ rounds on a (nontrivial) restricted class of network graphs.

Towards deriving these improved distributed algorithms, our main contribution is a general transformation that simulates any work-efficient PRAM algorithm running in $T$ parallel rounds via a distributed algorithm running in $T\cdot \tmix(G)\cdot 2^{O(\sqrt{\log n})}$ rounds.
Work- and time-efficient parallel algorithms for all of the aforementioned problems follow by combining the work of Sherman [FOCS'13, SODA'17] and Peng and Spielman [STOC'14]. Thus, simulating these parallel algorithms using our transformation framework produces the desired distributed algorithms.

The core technical component of our transformation is the algorithmic problem of solving \emph{multi-commodity routing}---that is, roughly, routing $n$ packets each from a given source to a given destination---in random graphs. For this problem, we obtain a new algorithm running in $2^{O(\sqrt{\log n})}$ rounds, improving on the $2^{O(\sqrt{\log n \log\log n})}$ round algorithm of Ghaffari, Kuhn, and Su [PODC'17]. As a consequence, for the MST problem in particular, we obtain an improved distributed algorithm running in $\tmix(G)\cdot 2^{O(\sqrt{\log n})}$ rounds.
\end{abstract}


\section{Introduction and Related Work}
This paper presents a general method that allows us to transform work-efficient parallel algorithms---formally in the PRAM model---into efficient distributed message-passing algorithms---formally in the CONGEST model---for a wide range of network graphs of practical interest. We believe that this method can be of significance for the following reasons: (1) parallel algorithms have been studied extensively since the late 1970s~\cite{fortune1978parallelism, goldschlager1978unified, savitch1979time} and there is a vast collection of known parallel algorithms for a variety of problems, and (2) there is a rather active community of research on developing new parallel algorithms. 
Our transformation opens the road for exporting these algorithms to the distributed setting and bridging the research in these two subareas in a concrete and formal manner. As immediate corollaries, by translating the recent work-efficient parallel algorithms for flow-type problems, we obtain new distributed algorithms for approximate maximum flow, shortest path, and transshipment.

Of course, such a transformation is bound to have some limitations. Due to the reasons that shall be explained soon, such a general transformation would be inefficient in worst-case network graphs. But we show that there are efficient transformations for many graph families of practical interest, and we also exhibit that these transformations entail interesting and non-trivial theoretical aspects.  

To explain our transformations, we first recall (informal) descriptions of the two computational models that we discuss, the distributed model and the parallel model. The more detailed model definitions are presented later in \Cref{sec:parallel-to-Dist}.

\paragraph{The Distributed Computing Model---CONGEST~\cite{Peleg:2000}:}
The network is abstracted as an $n$-node undirected graph $G=(V, E)$. There is one processor on each node of the network. At the risk of a slight informality, we use the words processor and node interchangeably. Each node has a unique $\Th(\log n)$-bit identifier. Communication happens in synchronous rounds where per round, each node can send one $B$-bit message to each of its neighboring nodes in the network graph $G$, where typically one assumes $B=O(\log n)$. During each round, each processor can perform unbounded computation with the information that it has at the time. The graph is known in a distributed fashion: each processor knows the edges incident on its own node. In case that the edges are weighted, the weight is known to both endpoints. At the end of the computation, each node should know its own part of the output: e.g., in computing a coloring, each node should know its own color.
One can measure the efficiency of an algorithm in the CONGEST model in different ways, such as number of rounds taken, or total number of messages sent among all nodes. In this paper, we only focus on minimizing the number of rounds that an algorithm takes.

\paragraph{The Parallel Model---PRAM~\cite{karp1988survey, kumar1994introduction}:} The system is composed of $p$ processors, each with a unique ID in $\{1, 2, \dots, p\}$, and a shared memory block of $M$ entries, including an output tape. In every round, each processor can read from or write to any memory entry (Concurrent Read and Concurrent Write, aka, CRCW); if multiple processors write to the same entry, an arbitrary one takes effect.\footnote{We can also support parallel algorithms that work under a more powerful model: if multiple processors write to the same memory, then we can take any associative function (min, max, sum) on the words written, and write that result into memory. However, for simplicity, we will work under the arbitrary CRCW model.} The input is provided in the shared memory cells in a manner that can be addressed easily, e.g., in the case of a graph, the input can be given as an adjacency list where there is one memory cell for the $j^{th}$ neighbor of the $i^{th}$ node.

\paragraph{Limitations to General Transformations?} Notice that the two models are intrinsically focused on different issues. The PRAM model is about speeding up computations, via using more processors, and tries to understand how much parallelism can help in reducing time. On the other hand, the distributed model is relevant where the system is by nature made of autonomous entities, each of which knows a part of the problem. For instance, in computer networks, which were historically the primary motivation for distributed models such as CONGEST, the computers in the network each know a part of the network graph and they cooperate to compute something about it, e.g., variants  of shortest paths or routing tables. Here, \emph{locality of the data} and \emph{limited communication bandwidth} are the main challenges. As such, it is arguably unreasonable to seek a general efficient transformation of \emph{any} parallel algorithm to a distributed one in \emph{any} arbitrary network graph. Let us elaborate on this. (1) The PRAM model is not limited by any \emph{locality}---each processor can asses any single register---while this is an intrinsic limitation in distributed systems---it can take time proportional to the diameter of the network graph for a processor to be informed of some bit residing in a far away corner of the network graph \footnote{And that bit may be relevant, as is in global problems such as minimum spanning tree, shortest path, etc.}. (2) Similarly, the network graph may have a small cut, which means transferring information across this cut, i.e., from the processors on one side of the cut to the other size, may take a long time, while this can be done much faster in the PRAM model. 

\paragraph{So What Can We Hope For?}
The above discussions and the two concrete points on \emph{locality} and \emph{congestion} (or in other words communication bandwidth) suggest that there may be some hope left: at least in network graphs that satisfy some mild conditions on diameter and cut sizes (or alternatively expansion, conductance, or other forms of formalizing lack of ``communication bottlenecks''), we might be able to find some general transformation. Arguably, these actually capture a range of network graphs of practical interest. For instance, overlay and peer-to-peer networks are designed and dynamically maintained over time in a fashion that ensures these good properties. 

One way of classifying some such nice graph families is by selecting all graphs whose mixing time for a random walk is relatively small. We define mixing time in Section~\ref{sec:transformation}, but informally, the mixing time of a graph is the number of steps a lazy random walk needs to take so that the distribution of the last vertex of the walk is roughly uniform over all $n$ vertices.
A wide range of the (overlay) networks used in practical distributed applications exhibit a good (e.g. polylogarithmic in $n$) mixing time. This holds for example for the networks in consideration in \cite{augustine2015enabling, awerbuch2004hyperring, mahlmann2005P2P, pandurangan2014dex, expander2003distributed, pandurangan2003P2P, pandurangan2011xheal, stoica2001chord}. 

A canonical reason for this good mixing time is because many of these overlay networks are formed in a way where each node is connected to $\Theta(\log n)$ randomly chosen nodes.  
Indeed, we present our general transformation primarily for such random graphs. We then also explain how to emulate the communication on random graphs atop arbitrary networks with a round-complexity overhead related to the mixing time of the graph, thus enabling us to extend the transformation to general graphs, with a round complexity overhead proportional to the mixing time.

\subsection{Our Results}
Our results build off of those in~\cite{ghaffari2017distributed}, whose main result is a distributed MST problem running in nearly \textit{mixing time}. We improve upon their results in two dimensions, one technical and one primarily conceptual. The technical contribution is an improved algorithm for the \textit{multicommodity routing} problem in random graphs, which is equivalent to the \emph{permutation routing} problem in~\cite{ghaffari2017distributed} up to $\tO(1)$ factors. We solve this problem in $2^{O(\srlogn)}$ rounds, improving upon the $2^{O(\sr{\logn\log\logn})}$ round algorithm in~\cite{ghaffari2017distributed}. Together with the ideas in~\cite{ghaffari2017distributed}, this immediately improves the distributed MST algorithm from $\tmix(G)\cd 2^{O(\sqrt{\log n\log\log n})}$ to $\tmix(G)\cd 2^{O(\srlogn)}$.

Our second, more conceptual contribution is in applying the multicommodity routing problem in a more general way. In particular, we use it to develop a framework that work-efficient algorithms in the PRAM model to distributed algorithms. This \emph{transformation} allows us to port the recent work-efficient parallel algorithms~\cite{peng2014efficient,sherman2013nearly,sherman2017generalized,becker2016near} for approximate maximum flow, shortest path, and transshipment to run in the CONGEST model, taking $\tmix(G)\cd n^{o(1)}$ rounds for all three problems.

 We first describe our multi-commodity routing result for random graphs, our main technical result and a key component in our transformations. We believe that this multi-commodity routing scheme and the hierarchical graph partitioning underlying it may be of independent interest. We then state our transformation results and overview some of their applications in deriving efficient distributed algorithms for some central graph problems.

\subsubsection{Multicommodity Routing on Random Graphs}

\paragraph{Random Graph Model:} We work with the following random (multi-)graph model $G(n,d)$ is as follows: each node $v\in V$ picks $d=\Omega(\log n)$ random nodes in $V$ independently with replacement, called the \defn{outgoing} neighbors of $v$. The network graph consists of all edges $(u,v)$ where $u$ is an outgoing neighbor of $v$ or vice versa. For $d=\Om(\log n)$, this is equivalent with the Erd{\"o}s-R\'{e}nyi model $\m G(n,d/n)$~\cite{erdos1959random}, with high probability; we use our variant for convenience. \footnote{Moreover, for many other models of random graphs, we can embed one round of this model (i.e., connecting each node to $O(\log n)$ randomly selected nodes) with a small, typically $\poly(\log n)$ round, overhead. This would be by using $O(n\log n)$ random walks, $O(\log n)$ starting from each node, and walking them until the mixing time, which is like selected a random connection endpoint. This is similar to~\cite{ghaffari2017distributed}. In many random graph families, these walks would mix in $\poly(\log n)$ rounds~\cite{cooper2008random}.} 

\paragraph{Multicommodity Routing:} Consider a random graph $G(n,p)$ for $p=O(\log n)$, and suppose that we have pairs  of nodes $(s_i,t_i) \in V\times V$. Suppose each node $s_i$ 
wants to communicate with its respective node $t_i$; we assume that node $t_i$ does not know $s_i$ beforehand. Our goal is to identify a path $P_i$ in $G$ between each pair $s_i$ and $t_i$. We refer to this problem as \defn{multicommodity routing}, to be formally defined in Section~\ref{sec:Routing}. In addition, if every node $v\in V$ appears at most $W$ times as $s_i$ or $t_i$, then we say that this multicommodity routing instance has \defn{width} $W$.

Our main technical contribution is an improved multi-commodity routing algorithm on random graphs with round complexity $2^{O(\srlogn)}$. This improves
on a solution of Ghaffari et al.~\cite{ghaffari2017distributed} which has round complexity $2^{O(\sr{\logn\log\logn})}$.

\begin{restatable}{theorem}{RoutingRandom}\label{thm:RoutingRandom}
Consider a multicommodity routing instance of width $\tO(1)$.
There is a multicommodity routing algorithm on $G(n,\Om(\log n))$ that achieves congestion and dilation $2^{O(\srlogn)}$, and runs in time $2^{O(\srlogn)}$.
\end{restatable}

\para{General Graphs and Mixing Time:} In fact, our result generalizes to more than random graphs in the same way as~\cite{ghaffari2017distributed}. As shown by ~\cite{ghaffari2017distributed}, random graphs can be ``embedded'' into any network graph with an overhead proportional to the \textit{mixing time} $\tmix$ of the network graph, which we define below. Thus, we can generalize the multicommodity routing algorithm to work on any graph.

Identically to~\cite{ghaffari2017distributed}, we define (lazy) random walks as follows: in every step, the walk remains at the current node with probability $1/2$, and otherwise, it transitions to a uniformly random neighbor. We formally define the mixing time of a graph as follows:
\BD
For a node $u\in V$, let $\{P_u^t(v)\}_{v\in V}$ be the probability distribution on the nodes $v\in V$ after $t$ steps of a (lazy) random walk starting at $u$. The \defn{mixing time} of the graph, denoted $\tmix$, is the minimum integer $t$ such that for all $u,v\in V$, $\left|P_u^t(v)-\frac{\deg(v)}{2m}\right|\le\frac{\deg(v)}{2mn}$.
\ED

Our multicommodity routing algorithm for general graphs is therefore as follows:

\begin{restatable}{theorem}{Main}\label{thm:Main}
There is a distributed algorithm solving multicommodity routing in $\tmix\cd2^{O(\srlogn)}$ rounds.
\end{restatable}

Finally, by substituting our multicommodity routing algorithm into the one in~\cite{ghaffari2017distributed}, we get an improvement on distributed MST in mixing time.
\BT
There is a distributed MST algorithm running in $\tmix\cd2^{O(\srlogn)}$ rounds.
\ET


\subsubsection{Transformation}\label{sec:transformation}
Our second, more conceptual contribution is a transformation from parallel algorithms to distributed algorithms on random graphs. In particular, we show that any work-efficient parallel algorithm running in $T$ rounds can be simulated on a distributed random graph network in $T\cd\tmix\cd2^{O(\srlogn)}$ rounds. The actual theorem statement, \Cref{thm:transform}, requires formalizing the parallel and distributed models, so we do not state it here.

\para{Applications:}
For applications of this transformation, we look at a recent line of work on near-linear time algorithms for flow-type problems. In particular, we investigate the approximate versions of shortest path, maximum flow, and transshipment (also known as uncapacitated minimum cost flow). Parallel $(1\pm\e)$-approximation algorithms for these problems running in $O(m^{1+o(1)})$ work and $O(m^{o(1)})$ time result from gradient descent methods combined with a parallel solver for symmetric diagonally dominant systems~\cite{peng2014efficient,sherman2013nearly,sherman2017generalized,becker2016near}. Therefore, by combining these parallel algorithms with our distributed transformation, we obtain the following corollaries:

\begin{restatable}{corollary}{Flow}
There are distributed algorithms running in time $\tmix \cd 2^{O(\srlogn)}$ for $(1+\e)$-approximate single-source shortest path and transshipment, and running time $\tmix \cd 2^{O(\sqrt{\logn\log\logn})}$ for $(1-\e)$-approximate maximum flow.
\end{restatable}
Finally, in the case of random graphs, another classical problem is the computation of a Hamiltonian cycle. Since an $\tO(n)$-work, $\tO(1)$-time parallel algorithm is known~\cite{coppersmith1987parallel}, we have an efficient distributed algorithm to compute Hamiltonian cycles.
\begin{restatable}{corollary}{Hamilton}
For large enough constant $C$, we can find a Hamilton cycle on $G(n,d)$ with $d=C\log n$ in $2^{O(\srlogn)}$ rounds, w.h.p.
\end{restatable}


This problem has attracted recent attention in the distributed setting. The main result of~\cite{chatterjee2018fast} is a distributed Hamiltonian cycle algorithm that runs in $\Om(n^{\de})$ rounds for graphs $G(n,d)$ with $d=\Om(\log n/n^\de)$ for any constant $0<\de\le1$. Thus, our algorithm greatly improves upon their result, both in number of rounds and in the parameter $d$.

\subsection{Some Other Related Work}

There has been a long history~\cite{valiant1990bridging,culler1993logp,karp1996efficient,pietracaprina1997complexity,fantozzi2003general} in translating the ideal PRAM model into more practical parallel models, such as the celebrated BSP model of Valiant~\cite{valiant1990bridging}. These transformations typically track many more parameters, such as communication and computation, than our transformation from PRAM to CONGEST, which only concerns the round complexity of the CONGEST algorithm.

There has also been work in the intersection of distributed computing and algorithms on random graphs. The task of computing a Hamiltonian cycle on a random graph was initiated by Levy et al.~\cite{levy2004distributed} and improved recently in~\cite{chatterjee2018fast}. Computation of other graph-theoretic properties on random graphs, such as approximate minimum dominating set and maximum matching, has been studied in a distributed setting in~\cite{krzywdzinski2015distributed}.

\section{Multicommodity Routing}\label{sec:Routing}

We formally define the multicommodity routing problem below, along with the congestion and dilation of a solution to this problem.

\BD
A multicommodity routing instance consists of pairs of nodes $(s_i,t_i)\in V\times V$, such that each $t_i$ is known to node $s_i$. A solution consists of a (not necessarily simple) path $P_i$ connecting nodes $s_i$ and $t_i$ for every $i$, such that every node on $P_i$ knows its two neighbors on $P_i$.

The input has \defn{width} $W$ if every node $v\in V$ appears at most $W$ times as $s_i$ or $t_i$.

For a given solution of paths, the \defn{dilation} is the maximum length of a path, and the \defn{congestion} is the maximum number of times any edge appears in total over all paths. More precisely, if $c_i(e)$ is the number of occurrences of edge $e\in E(G)$ in path $P_i$, then the congestion is $\max_{e\in E(G)}\sum_ic_i(e)$.
\ED

The significance of the congestion and dilation parameters lies in the following lemma from~\cite{ghaffari2016distributed}, whose proof uses the standard trick of \textit{random delays} from packet routing~\cite{leighton1988universal}. In particular, if a multicommodity routing algorithm runs efficiently and outputs a solution of low congestion and dilation, then each node $s_i$ can efficiently route messages to node $t_i$.

\BT[\cite{ghaffari2016distributed}]\label{thm:CongestionDilation}
Suppose we solve a multicommodity routing instance $\{(s_i,t_i)\}_i$ and achieve congestion $c$ and dilation $d$. Then, in $\tO(c+d)$ rounds, every node $s_i$ can send one $O(\log n)$-bit message to every node $t_i$, and vice versa.
\ET

We now provide our algorithm for multicommodity routing, improving the congestion and dilation factors from $2^{O(\sr{\logn\log\logn})}$ in~\cite{ghaffari2017distributed} to $2^{O(\srlogn)}$. Like~\cite{ghaffari2017distributed}, our algorithm uses the concept of \emph{embedding} a graph, defined below.

\BD
Let $H$ and $G$ be two graphs on the same node set. We say that an algorithm \defn{embeds} $H$ into $G$ with congestion $c$ and dilation $d$ if the algorithm solves the following multicommodity routing instance on $G$: the $(s_i,t_i)$ pairs are precisely the edges of $H$, the congestion is $c$, and the dilation is $d$.
For each $(s,t)\in E(H)$, the path $P_{s,t}$ (in $G$ from $s$ to $t$) is called the \defn{embedded path} for edge $(s,t)$. 
\ED

Our multicommodity routing algorithm will \emph{recursively} embed graphs. We use the following helper lemma.

\BL\label{lem:Embedding}
Suppose there is a distributed algorithm $\m A_1$ embedding graph $G_1$ into network $G_0$ with congestion $c_1$ and dilation $d_1$ in $T_1$ rounds, and another distributed algorithm $\m A_2$ embedding graph $G_2$ into network $G_1$ with congestion $c_2$ and dilation $d_2$ in $T_2$ rounds. Then, there is a distributed algorithm embedding $G_2$ into network $G_0$ with congestion $c_1c_2$ and dilation $d_1d_2$ in $T_1+T_2\cd\tO(c_1+d_1)$ rounds.
\EL
\BP
First, we provide the embedding without the algorithm. For each pair $(s,t)\in E(G_1)$, let $P^1_{s,t}$ be the embedded path in $G_0$, and for each pair $(s,t)\in E(G_2)$, let $P^2_{s,t}$ be the embedded path in $G_1$. To embed edge $(s,t)\in E(G_2)$ into $E_0$, consider the path $P^2_{s,t}:=(s=v_0,v_1,v_2,\lds,v_{\el}=t)$; the embedded path for $(s,t)$ in $G_0$ is precisely the concatenation of the paths $P^1_{v_{i-1},v_{i}}$ for $i\in[\el]$ in increasing order. Since $\el\le d_2$ and each path $P^1_{v_{i-1},v_{i}}$ has length at most $d_1$, the total length of the embedded path for $(s,t)$ in $G_0$ is at most $d_1d_2$, achieving the promised dilation.

For congestion, let $c^1_{s,t}(e)$ denote the number of occurrences of edge $e\in E(G_0)$ in $P^1_{s,t}$. Since each edge $(s,t)\in E(G_1)$ shows up at most $c_2$ times among all $P^2_{s',t'}$, the number of times the path $P^1_{s,t}$ is concatenated in the embedding is at most $c^1_{s,t}(e)\cd c_2$. Therefore, edge $e\in E(G_0)$ occurs at most $\sum_{s,t}c^1_{s,t}(e)\cd c_2\le c_1c_2$ times among all the concatenated paths embedding $G_2$ into $G_0$.

Finally, we describe the embedding algorithm. First, the algorithm on $G_0$ runs $\m A_1$, obtaining the embedding of $G_1$ into $G_0$ in $T_1$ rounds. We now show how to emulate a single round of $\m A_2$ running on network $G_1$ using $\tO(c_1+d_1)$ rounds on network $G_0$. Suppose that, on a particular round, $\m A_2$ has each node $s$ send a message $x$ to node $t$ for every $(s,t)\in E(G_1)$. Since the embedding of $G_1$ into $G_0$ is a multicommodity routing instance, we use Theorem~\ref{thm:CongestionDilation}, where each node $s$ tries to route that same message $x$ to node $t$. This runs in $\tO(c_1+d_1)$ rounds for a given round of $\m A_2$. Altogether, we spend $T_1+T_2\cd\tO(c_1+d_1)$ rounds to emulate the entire $\m A_2$.
\EP

We now prove Theorem~\ref{thm:RoutingRandom}, restated below.

\RoutingRandom*

\BP
Following \cite{ghaffari2017distributed}, our strategy is to construct graph embeddings recursively, forming a hierarchical decomposition. We start off by embedding a graph of sufficiently high degree in $G$, similar to the ``Level Zero Random Graph'' embedding of~\cite{ghaffari2017distributed}. Essentially, the embedded paths are random walks in $G$ of length $\tmix$; we refer the reader to~\cite{ghaffari2017distributed} for details.

\BL[\cite{ghaffari2017distributed}, Section 3.1.1]\label{lem:LevelZero}
On any graph $G$ with $n$ nodes and $m$ edges, we can embed a random graph $G(m,d)$ with $d\ge200\logn$ into $G$ with congestion $\tO(\tmix\cd d)$ and dilation $\tmix$ in time $\tO(\tmix \cd d)$.
\EL

For our instance, $\tmix=O(\log n)$ since $G\sim G(n,\Om(\log n))$. Let $d:=2^{10\srlogn}$. For this value of $d$ in the lemma, we obtain an embedding $G_0\sim G(m,d)$ into $G$ in time $\tO(2^{10\srlogn})$. 

Similarly to~\cite{ghaffari2017distributed}, our first goal is to obtain graphs $G_1,G_2,\lds,G_K$ which form some hierarchical structure, such that each graph $G_i$ embeds into $G_{i-1}$ with small congestion and dilation. Later on, we will exploit the hierarchical structure of the graphs $G_0,G_1,G_2,\lds,G_K$ in order to route each $(s_i,t_i)$ pair.

To begin, we first describe the embedding of $G_1$ into $G_0$. Like \cite{ghaffari2017distributed}, we first randomly partition the nodes of $G_0$ into $\be$ sets $A_1,\lds,A_\be$ so that $|A_i|=\Th(m/\be)$. Our goal is to construct and embed $G_1$ into $G_0$ with congestion $1$ and dilation $2$, where $G_1$ has the following structure: it is a disjoint union, over all $i\in[\be]$, of a random graph $G_1^{(i)}\sim G(|A_i|,d/4)$ on the set $A_i$. By definition, $G_0$ and $G_1$ share the same node set.  Note that \cite{ghaffari2017distributed} does a similar graph embedding, except with congestion and dilation $O(\logn)$; improving the factors to $O(1)$ is what constitutes our improvement.

Fix a set $A_i$; we proceed to construct the random graph in $A_i$. For a fixed node $u\in V(G_0)$, consider the list of outgoing neighbors of $u$ in $A_i$. Note that since $G_0$ can have multi-edges, a node in $A_i$ may appear multiple times in the list. Now, inside the local computation of node $u$, randomly group the nodes in the list into ordered pairs, leaving one element out if the list size is odd. For each ordered pair $(v_1,v_2)\in A_i\times A_i$, add $v_2$ into $v_1$'s list of outgoing edges in $G_1^{(i)}$, and embed this edge along the path $(v_1,u,v_2)$ of length $2$. In this case, since the paths are short, node $u$ can inform each pair $(v_1,v_2)$ the entire path $(v_1,u,v_2)$ in $O(1)$ rounds.

Since node $u$ has $d$ outgoing neighbors, the expected number of outgoing neighbors of $u$ in $A_i$ is $d/\be$. By Chernoff bound, the actual number is at least $0.9d/\be$ w.h.p., so there are at least $0.4d/\be$ ordered pairs w.h.p. Over all nodes $u$, there are at least $0.4md/\be$ pairs total.

We now argue that, over the randomness of the construction of $G_0$, the pairs are uniformly and independently distributed in $A_i\times A_i$. We show this by revealing the randomness of $G_0$ in two steps.
If, for each node $u$, we first reveal which set $A_j$ each outgoing neighbor of $u$ belongs to, and then group the outgoing neighbors in $A_i$ into pairs, and finally reveal the actual outgoing neighbors, then each of the at least $0.4md/\be$ pairs is uniformly and independently distributed in $A_i\times A_i$. Therefore, each node $v\in A_i$ is expected to receive at least $ \f{0.4md/\be}{m/\be}=0.4d$ outgoing neighbors, or at least $ 0.25d$ outgoing neighbors w.h.p.\ by Chernoff bound. Finally, we have each node in $A_i$ randomly discard outgoing neighbors until it has $d/4$ remaining. The edges remaining in $A_i$ form the graph $G_1^{(i)}$, which has distribution $G(|A_i|,d/4)$. Thus, we have embedded a graph $G_1$ consisting of $\be$ disjoint random graphs $G(\Th(m/\be), d/4)$ into $G_0$ with congestion $1$ and dilation $2$.

We apply recursion in the same manner as in \cite{ghaffari2017distributed}: recurse on each $G_1^{(i)}$ (in parallel) by partitioning its vertices into another $\be$ sets $A_1,\lds,A_\be$, building a random graph on each set, and taking their disjoint union.
More precisely, suppose the algorithm begins with a graph $H_0\sim G(|V(G')|,d/4^{t-1})$ on depth $k$ of the recursion tree (where the initial embedding of $G_1$ into $G_0$ has depth $1$). The algorithm randomly partitions the nodes of $H$ into $A_1,\lds,A_\be$ and defines a graph $H_1$ similar to $G_1$ from before: it is a disjoint union, over all $i\in[\be]$, of a random graph $H_1^{(i)}\sim G(|A_i|,d/4^t)$ on the set $A_i$. Finally, the algorithm recurses on each $H_1^{(i)}$. This recursion stops when the graphs have  size at most $2^{5\srlogn}$; in other words, if $|V(H_0)|\le2^{5\srlogn}$, then the recursive algorithm exits immediately instead of performing the above routine.

Once the recursive algorithm finishes, we let $G_k$ be the disjoint union of all graphs $H^{(i)}_1$ constructed on a recursive call of depth $k$. Observe that $G_k$ has the same node set as $G_0$. Moreover, since, on each recursive step the sizes of the $A_i$ drop by a factor of $1/\be$ in expectation, or at most $ 2/\be$ w.h.p., the recursion goes for at most $\log_{\be/2}n \le 2\srlogn$ levels. Therefore, for each disjoint random graph in each $G_k$, the number of outgoing neighbors is always at least $d/4^{2\srlogn}\ge 2^{6\srlogn}$. In addition, since every embedding of $G_{k}$ into $G_{k-1}$ has congestion $1$ and dilation $2$, by applying Lemma~\ref{lem:Embedding} repeatedly, $G_K$ embeds into $G_0$ with congestion $1$ and dilation $2^{2\srlogn}$, and into $G$ with congestion and dilation $2^{O(\srlogn)}$. Moreover, on each recursion level $k$, the embedding algorithm takes a constant number of rounds on the graph $G_{k-1}$, which can be simulated on $G$ in $2^{O(\srlogn)}$ rounds by Lemma~\ref{lem:Embedding}.

Now we discuss how to route each $(s_i,t_i)$ pair. Fix a pair $(s,t)$; at a high level, we will iterate over the graphs $G_0,G_1,G_2,\lds$ while maintaining the invariant that $s$ and $t$ belong to the same connected component in $G_k$. Initially, this holds for $G_0$; if it becomes false when transitioning from $G_{k-1}$ to $G_k$, then we replace $s$ with a node $s'$ in the connected component of $t$ in $G_k$. We claim that in fact, w.h.p., there is such a node $s'$ that is \emph{adjacent} to $s$ in $G_{k-1}$; hence, $s$ can send its message to $s'$ along the network $G_{k-1}$, and the algorithm proceeds to $G_k$ pretending that $s$ is now $s'$.
 This process is similar to that in \cite{ghaffari2017distributed}, except we make do without their notion of ``portals'' because of the large degree of $G_0$---$2^{\Th(\srlogn)}$ compared to $\Th(\logn)$ in \cite{ghaffari2017distributed}. 

We now make the routing procedure precise. For a given $G_k$ with $k<K$, if $s$ and $t$ belong to the same connected component of $G_k$, then we do nothing. Otherwise, since $s$ has at least $2^{6\srlogn}=\om(\be\logn)$ neighbors, w.h.p., node $s$ has an outgoing neighbor $s'$ in the connected component of $G_k$ containing $t$; if there are multiple neighbors, one is chosen at random. Node $s$ \emph{relays} the message along this edge to $s'$, and the pair $(s,t)$ is replaced with $(s',t)$ upon applying recursion to the next level. \footnote{In reality, node $s$ does not know which set $A_i$ contains node $t$. Like~\cite{ghaffari2017distributed}, we resolve this issue using $\tO(1)$-wise independence, which does not affect the algorithm's performance. Since $\Th(W\logn)$ bits of randomness suffice for $W$-wise independence~\cite{Alon-Spencer}, we can have one node draw $\Th(W\log n)=\tO(1)$ random bits at the beginning of the iteration and broadcast them to all the nodes in $\tO(1)$ time. Then, every node can locally compute the set $A_i$ that contains any given node $t$; see~\cite{ghaffari2017distributed} for details.} Therefore, we always maintain the invariant that in each current $(s,t)$ pair, both $s$ and $t$ belong in the same random graph. 

We now argue that w.h.p., each vertex $s'$ has $\tO(1)$ messages after this routing step. By assumption, every node $v\in V$ appears $\tO(1)$ times as $t_j$, so there are $|A_i|\cd\tO(1)$ many nodes $t_j$ that are inside $A_i$. For each such $t_j$ with $s_j\notin A_i$, over the randomness of $G_{k-1}$, the neighbor $s'_j$ of $s_j$ inside $A_i$ chosen to relay the message from $s_j$ is uniformly distributed in $A_i$. By Chernoff bound, each node in $A_i$ is chosen to relay a message $\tO(1)$ times when transitioning from $G_{k-1}$ to $G_k$. In total, each node $v\in V$ appears $\tO(1)$ times as $s_i$ in the beginning, and receives $\tO(1)$ messages to relay for each of $O(\srlogn)$ iterations. It follows that every node always has $\tO(1)$ messages throughout the algorithm.

Finally, in the graph $G_K$, we know that each $(s_j,t_j)$ pair is in the same connected component of $G_K$. Recall that each connected component in $G_K$ has at most $2^{5\srlogn}$ nodes, each with degree at least $2^{6\srlogn}$ (possibly with self-loops and parallel edges). It follows that w.h.p., each connected component is a ``complete'' graph, in the sense that every two nodes in the component are connected by at least one edge. Therefore, we can route each $(s_j,t_j)$ pair trivially along an edge connecting them. 

As for running time, since each graph  $G_0,G_1,\lds,G_K$ embeds into $G$ with congestion and dilation $2^{O(\srlogn)}$ by Lemma~\ref{lem:Embedding}, iterating on each graph $G_k$ takes $2^{O(\srlogn)}$ rounds. Therefore, the total running time is $2^{O(\srlogn)}$, concluding Theorem~\ref{thm:RoutingRandom}.
\EP

For general graphs, we can repeat the same algorithm, except we embed $G_0$ with congestion and dilation $\tO(\tmix\cd2^{10\srlogn)})$ instead of $\tO(2^{10\srlogn})$, obtaining the following:

\BC\label{thm:RoutingGeneral}
Consider a multicommodity routing algorithm where every node $v\in V$ appears $\tO(1)$ times as $s_i$ or $t_i$.
There is a multicommodity routing algorithm that achieves congestion and dilation $\tmix\cd2^{O(\srlogn)}$, and runs in time $\tmix\cd2^{O(\srlogn)}$.
\EC

Combining Theorem~\ref{thm:CongestionDilation} and Corollary~\ref{thm:RoutingGeneral} proves Theorem~\ref{thm:Main}.

\section{Parallel to Distributed}
\label{sec:parallel-to-Dist}
In this section, we present our procedure to simulate parallel algorithms on distributed graph networks.

\para{Parallel Model Assumptions.} To formalize our transformation, we make some standard input assumptions to work-efficient parallel algorithms:


\BE
\im The input graph is represented in adjacency list form. There is a pointer array of size $n$, whose $i$'th element points to an array of neighbors of vertex $v_i$. The $i$'th array of input begins with $\deg(v_i)$, followed by the $\deg(v_i)$ neighbors of vertex $v_i$.
\im There are exactly $2m$ processors.\footnote{If the algorithm uses $m^{1+o(1)}$ processors, then we can have each of the $2m$ processors simulate $m^{o(1)}$ of them.} 
Each processor knows its ID, a unique number in $[n]$, and has unlimited local computation and memory.
\im There is a shared memory block of $\tO(mT)$ entries, including the output tape,  where $T$ is the running time of the parallel algorithm.\footnote{If the algorithm uses much more than $mT$ memory addresses,  then we can hash the memory addresses down to a hash table of $\tO(mT)$ entries.} In every round, each processor can read or write from any entry in unit time (CRCW model). If multiple processors write to the same entry on the same round, then an arbitrary write is selected for that round.
\im If the output is a subgraph, then the output tape is an array of the subgraph edges.
\EE
\para{Distributed Model Assumptions.}Similarly, we make the following assumptions on the distributed model.
\BE
\im Each node knows its neighbors in the input graph, as well as its ID, a unique number of $\Th(\log n)$ bits. Each node has unlimited local computation and memory.
\im If the output is a subgraph, each node should know its incident edges in the subgraph.
\EE

\begin{restatable}{theorem}{Transform}\label{thm:transform}
Under the above parallel and distributed model assumptions, a parallel graph algorithm running in $T$ rounds can be simulated by a distributed algorithm in $T\cd\tmix\cd2^{O(\srlogn)}$ rounds.
\end{restatable}

\BP
We want to simulate one round of the parallel algorithm in $\tmix\cd2^{O(\srlogn)}$ rounds in the distributed model. To do so, we need to simulate the processors, input data, shared memory, and output.

\para{Processors.}
Embed a random graph $G_0=G(2m,\Th(\log n/m))$ into the network graph, as in \cite{ghaffari2017distributed}. Every node in $G_0$ simulates one processor so that all $2m$ processors are simulated; this means that every node in the original network simulates $\deg(m)$ processors. Let the nodes of $G_0$ and the processors be named $(v,j)$, where $v\in V$ and $j\in[\deg(v)]$. Node/processor $(v,j)$ knows the $j$'th neighbor of $v$, and say, $(v,1)$ also knows the value of $\deg(i)$. Therefore, all input data to the parallel algorithm is spread over the processors $(v,j)$. From now on, we treat graph $G_0$ as the new network graph in the distributed setting.

\para{Shared memory.}
Shared memory is spread over all $2m$ processors. Let the shared memory be split into $2m$ blocks of size $B$ each, where $B:=\tO(1)$. Processor $(v_i,j)$ is in charge of block $\sum_{i'<i}\deg(v_{i'})+j$, so that each block is maintained by one processor. To look up block $k$ in the shared memory array, a processor needs to write $k$ as $\sum_{i'<i}\deg(v_{i'})+j$ for some $(v_i,j)$. Suppose for now that each processor knows the map $\phi:[2m]\to V\times \N$ from index $k$ to tuple $(v_i,j)$; later on, we remove this assumption. 

On a given parallel round, if a processor asks for block $k$ of shared memory, it sends a request to node $\phi(k)$. One issue is the possibility that many nodes all want to communicate with processor $\phi(k)$, and in the multicommodity routing problem, we only allow each target node to appear $\tO(1)$ times in the $(s_i,t_i)$ pairs. We solve this issue below, whose proof is deferred to Appendix~\ref{sec:HighDegree}.

\begin{restatable}{lemma}{VirtualTree}\label{lem:VirtualTree}
Consider the following setting: there is a node $v_0$, called the root, in possession of a memory block, and nodes $v_1,\lds,v_k$, called leaves, that request this memory block. The root node does not know the identities of the leaf nodes, but the leaf nodes know the identity $v_0$ of the root node. Then, in $\tO(1)$ multicommodity routing calls of width $\tO(1)$, the nodes $v_1,\lds,v_k$ can receive the memory block of node $v_0$.

Now consider multiple such settings in parallel, where every node in the graph is a root node in at most one setting, and a leaf node in at most one setting. Then, in $\tO(1)$ multicommodity routing calls of width $\tO(1)$, every leaf node can receive the memory block of its corresponding root node.
\end{restatable}

To remove the assumption that each processor knows the map $\phi$, we do the following as a precomputation step. We allocate an auxiliary array of size $n$, and our goal is to fill entry $i$ with $\sum_{i'<i}\deg(v_{i'})+j$. Let processor $(v_i,1)$ be in charge of entry $i$. Initially, processor $(v_i,1)$ fills entry $i$ with $\deg(v_i)$, which it knows. Then, getting the array we desire amounts to computing prefix sums, and we can make the parallel prefix sum algorithm work here~\cite{ladner1980parallel}, since any processor looking for entry $i$ knows to query $(v_i,1)$ for it. Finally, for a node to determine the entry $\phi(k)$, it can binary search on this auxiliary array to find the largest $i$ with $\sum_{i'<i}\deg(v_{i'})<k$, and set $j:=k-\sum_{i'<i}\deg(v_{i'})$, which is the correct $(v_i,j)$.

\para{Input data.}
If a processor in the parallel algorithm requests the value of $\deg(v)$ or the $i$'th neighbor of vertex $v$, we have the corresponding processor send a request to processor $(v,i)$ for this neighbor. The routing details are the same as above.

\para{Output.}
If the output is a subgraph of the original network graph $G$, then the distributed model requires each original node to know its incident edges in the subgraph. One way to do this is as follows: at the end of simulating the parallel algorithm, we can first sort the edges lexicographically using the distributed translation of a parallel sorting algorithm. Then, each node $(v_i,i)$ binary searches the output to determine if the edge of $v$ to its $i$'th neighbor $u$ is in the output (either as $(u,v)$ or as $(v,u)$). Since each original node $v\in V$ simulates each node/processor $(v_i,i)$, node $v$ knows all edges incident to it in the output subgraph.
\EP

\subsection{Applications to Parallel Algorithms}

The task of approximately solving symmetric diagonally dominant (SDD) systems $Mx=b$ appears in many fast algorithms for $\el_p$ minimization problems, such as maximum flow and transshipment. Peng and Spielman~\cite{peng2014efficient} obtained the first polylogarithmic time parallel SDD solver, stated below. For precise definitions of SDD, $\e$-approximate solution, and condition number, we refer the reader to~\cite{peng2014efficient}.

\BT[Peng and Spielman~\cite{peng2014efficient}]
The SDD system $Mx=b$, where $M$ is an $n\times n$ matrix with $m$ nonzero entries, can be $\e$-approximately solved in parallel in $\tO(m\log^3\kappa)$ work and $\tO(\log\kappa)$ time, where $\kappa$ is the condition number of matrix $M$.
\ET

Using our framework, we can translate this algorithm to a distributed setting, assuming that the input and output are distributed proportionally among the nodes.

\begin{restatable}{corollary}{SDD}
Let $G$ be a network matrix. Consider a SDD matrix $M$ with condition number $\kappa$, whose rows and columns indexed by $V$, and with nonzero entries only at entries $M_{u,v}$ with $(u,v)\in E$. Moreover, assume that each nonzero entry $M_{u,v}$ is known to both nodes $u$ and $v$, and that each entry $b_v$ is known to node $v$. In $\tO(\tmix\cd\log^4\kappa)$ distributed rounds, we can compute an $\e$-approximate solution $x$, such that each node $v$ knows entry $x_v$.
\end{restatable}

By combining parallel SDD solvers with gradient descent, we can compute approximate solutions maximum flow and minimum transshipment in parallel based on the recent work of Sherman and Becker et al.~\cite{sherman2013nearly,sherman2017generalized,becker2016near}. An added corollary is approximate shortest path, which can be reduced from transshipment~\cite{becker2016near}.

\BT[Sherman, Becker et al.~\cite{sherman2013nearly,sherman2017generalized,becker2016near}]
The $(1+\e)$-approximate single-source shortest path and minimum transshipment problems can be solved in parallel in $m\cd2^{O(\srlogn)}$ work and $2^{O(\srlogn)}$ time. The $(1-\e)$-approximate maximum flow problem can be solved in parallel in $m\cd2^{O(\sqrt{\logn\log\logn})}$ work and $2^{O(\sqrt{\logn\log\logn})}$ time.
\ET

\Flow*

Lastly, we consider the task of computing a Hamiltonian cycle on random graphs. This problem can be solved efficiently in parallel on random graphs $G(n,d)$, with $d=C\log n$ for large enough constant $C$, by a result of Coppersmith et al.~\cite{coppersmith1987parallel}. We remark that~\cite{coppersmith1987parallel} only states that their algorithm runs in $O(\log^2n)$ time \emph{in expectation}, but their proof is easily modified so that it holds w.h.p., at the cost of a larger constant $C$.

\BT[Coppersmith et al.~\cite{coppersmith1987parallel}]
For large enough constant $C$, there is a parallel algorithm that finds a Hamiltonian cycle in $G(n,C\log n)$ in $O(\log^2n)$ time, w.h.p.
\ET

This immediately implies our fast distributed algorithm for Hamiltonian cycle; the result is restated below.

\Hamilton*

\section{Conclusion and Open Problems}

In this paper, we bridge the gap between work-efficient parallel algorithms and distributed algorithms in the CONGEST model. Our main technical contribution lies in a distributed algorithm for multicommodity routing on random graphs.

The most obvious open problem is to improve the $2^{O(\srlogn)}$ bound in Theorem~\ref{thm:RoutingRandom}. Interestingly, finding a multicommodity routing solution with congestion and dilation $O(\log n)$ is fairly easy if we are allowed $\poly(n)$ time. In other words, while there exist good multicommodity routing solutions, we do not know how to find them efficiently in a distributed fashion. Hence, finding an algorithm that both runs in $\tO(1)$ rounds and computes a solution of congestion and dilation $\tO(1)$ is an intriguing open problem, and would serve as evidence that distributed computation on well-mixing network graphs is as easy as work-efficient parallel computation, up to $\tO(1)$ factors.

\bibliography{ref}

\appendix

\section{High Degree Communication}\label{sec:HighDegree}

\VirtualTree*

\BP (Lemma~\ref{lem:VirtualTree})
We assume that every node has a unique ID in the range $\{1,2,\lds,n\}$. The reduction from $\Th(\logn)$-bit identifiers is standard: construct a BFS tree of depth $D$, where $D$ is the diameter of the network graph, root the tree arbitrarily, and run prefix/infix/postfix ordering on the tree in $O(D)$ time. Since $\tmix\ge D$, this takes $O(\tmix)$ time, which is negligible.

For now, consider the first setting of the lemma, with only one root node. Our goal is to establish a low-degree and low-diameter tree of communication, which contains the leaf nodes and possibly other nodes. The root node can then send the memory block to one of the nodes in this tree, which then gets propagated to all other nodes on the tree, including the leaf nodes.

Let $K$ be a parameter that starts at $n/2$ and decreases by a factor of $2$ for $T:=\lc\log_2(n/2)\rc$ rounds. The node with ID $1$ picks a hash function $f:V\times [K]\to V$ for this iteration, and broadcasts it to all other nodes in $D$ rounds. At the end, we will address the problem of encoding hash functions, but for now, assume that  the hash function has mutual independence.  

On iteration $i$, each leaf node computes a private random number $k\in[K]$ and computes $f(v_0,k)\in V$, called the \textit{connection point} for leaf node $v_i$. We will later show that, w.h.p., each node in $V$ is the connection point of $\tO(1)$ leaf nodes. Assuming this, we form the multicommodity routing instance where each leaf node requests a routing to its connection point, so that afterwards, each connection point $v_j$ learns its set $S_j$ of corresponding leaf nodes. Each connection point elects a random node $v_j^*\in S_j$ as the \textit{leader}, and routes the entire set $S_j$ to node $v_j^*$ in another multicommodity routing instance. All nodes in $S_j\bs v_j^*$, which did not receive the set $S_j$, drop out of the algorithm, leaving the leader $v_j^*$ to route to other nodes in later iterations. At the end of the algorithm, there is only one leader left, and that leader routes directly to the root node $v_0$, receiving the memory block. Finally, the memory block gets propagated from the leaders $v_j^*$ to the other nodes in $S_j$ in reverse iteration order.

We now show that, w.h.p., each node in $V$ is a connection point to $\tO(1)$ leaf nodes; this would bound the width of the multicommodity instances by $\tO(1)$. Initially, there are at most $n$ leaf nodes and $n/2$ possible connection points, so each connection point has at most $2$ leaf nodes in expectation, or $O(\log n)$ w.h.p. On iteration $t>1$, there are at most $ n/2^{t-1}$ leaf nodes left, since each of the $n/2^{t-1}$ connection point elected one leader in the previous iteration and those are the only leaf nodes remaining. So each of the $n/2^t$ connection points has at most $2$ leaf nodes in expectation, or $O(\log n)$ w.h.p.

Now consider the general setting, where we do the same thing in parallel over all groups of leaf nodes. On iteration $t$, let the set of remaining leaf nodes in each setting be $L_1,\lds,L_r$. For each set of leaf nodes $L_i$, a given node $v_j$ has probability $1/2^t$ of being selected as a connection point for $L_i$, and if so, it is expected to have at most $\frac{|L_i|}{n/2^t}$ many leaf nodes in $L_i$, or $O(\frac{|L_i|}{n/2^t}\log n)=O(\log n)$ w.h.p., using that $|L_i|\le n/{2^{t-1}}$. Therefore, if $X_j^i$ is the random variable of the number of leaf nodes in $L_i$ assigned to node $v_j$, then $\E[X_j^i]\le|L_i|/n$, and $X_j^i=O(\log n)$ w.h.p. Conditioned on the w.h.p.\ statement, we use the following variant of Chernoff bound:
\BT[Chernoff bound]
If $X_1,\lds,X_n$ are independent random variables in the range $[0,C]$ and $\mu:=\E[X_1+\cds+X_n]$, then \[ \Pr[X_1+\cds+X_n\ge(1+\de)\mu] \le \exp\lp-\frac{2\de^2\mu^2}{nC^2}\rp .\]
\ET
Taking the independent variables $X_j^1,\lds,X_j^r$ and setting $\de:=\Th(\frac{\log^2n}\mu)$ and $C:=O(\log n)$, we get that $\mu=\sum_i|L_i|/n\le1$ and \[\Pr[X_j^1+\cds+X_j^r\ge\Th(\log^2n)]\le \exp(-O(\log^2n) ).\]
Therefore, w.h.p., every node has $O(\log^2n)$ neighbors at any given round.

Lastly, we address the issue of encoding hash functions, which we solve using $W$-wise independent hash families for a small value $W$. Since the algorithm runs in $\tO(1)$ rounds, $W=\tO(1)$ suffices. It turns out that deterministic families of $2^{O(W\log n)}$ hash functions exist~\cite{Alon-Spencer}, so the node with ID $1$ can simply pick a random $O(W\log n)=\tO(1)$-bit string and broadcast it to all other nodes in $D+\tO(1)=\tO(\tmix)$ rounds.
\EP

\end{document}